# Electrical resistance associated with the scattering of optically oriented electrons in n-GaAs

M.D. Ragoza[1], N.V. Kozyrev[1], S.V. Nekrasov[1], B.R. Namozov[1], Yu.G. Kusrayev[1], N. Bart[2], A. Ludwig[2], and A.D. Wieck[2]

[1]Ioffe Institute, Russian Academy of Sciences, 194021 St. Petersburg, Russia
[2]Lehrstuhl für Angewandte Festkörperphysik, Ruhr-Universität Bochum, Bochum, Germany

***Abstract.*** *In a bulk GaAs crystal, an unusual magnetoresistance effect, which takes place when a spin-polarized current flows through the sample, was detected. Under conditions of optical pumping of electron spins, an external magnetic field directed along the electric current and perpendicular to the oriented spins decreases the resistance of the material. The phenomenon is due to the spin-dependent scattering of electrons by neutral donors. It was found that the sign of the magnetoresistance does not depend on the sign of the exciting light circular polarization, the effect is even with respect to the sign of the spin polarization of the carriers, which indicates a correlation between the spins of optically oriented free electrons and electrons localized on donors.*

**Introduction.**

Spin-dependent interactions of electrons in semiconductors give rise to many important optical and transport effects, such as magnetoresistance, spin Hall effect, circular photovoltaic effect, optical orientation, spin relaxation, etc. [1]. The field of research on spin-dependent transport is very wide and extends from metals [2] to semiconductors [3] and, further, to organic materials [4].

The effects associated with spin-dependent transport can be used to control the spin of carriers used to transfer and process information in spintronic devices. A large and important group of these phenomena is united under the concept of magnetoresistance. The effect of magnetoresistance in the most general sense is understood as a change in the electrical resistance of a material under the action of an external magnetic field. One of the most well-known magnetoresistance mechanisms, negative magnetoresistance, is due to weak localization, which leads to an increase in conductivity when a magnetic field is applied [5, 6]. If one doesn't take into account the Lorentz force and the effects of weak localization, magnetoresistance is usually associated with spin-dependent carrier scattering or spin-dependent recombination. In metals, for example, the spin-dependent scattering of band electrons by localized spins leads to the Kondo effect [7].

An important role in the phenomena of spin-dependent transport in semiconductors is played by the exchange scattering of carriers [8]. Thus, scattering by a neutral donor type impurity, which often determines the mobility of charge carriers, is spin-dependent and resembles scattering of electrons on a hydrogen atom studied in atomic physics [9]. It was found that the electric resistance of an n-type silicon sample in a strong magnetic field depends on the spin polarization of free and donor-localized electrons due to spin-dependent recombination [10].

Spin dependent scattering also affects the rate of spin relaxation of charge carriers. It was theoretically and experimentally demonstrated in Ref. [11] that the inelastic scattering of conduction electrons by electrons bound to donors in silicon leads to a strong depolarization of their spins via spin exchange. This phenomenon reminds of spin relaxation due to the electron-hole exchange interaction, the so-called Bir-Aronov-Pikus mechanism [12]. In Refs. [13, 14], it was shown that in doped silicon devices the mechanism of exchange scattering is significantly superior to the Elliott-Yafet spin relaxation mechanism that dominates in other cases.

The theory of spin relaxation of carriers in semiconductors developed by Dyakonov and Perel [15, 16] is based on the spin-orbit interaction and is a weak relativistic effect. The exchange scattering process, on the contrary, is nonrelativistic, and even in materials with a weak spin-orbit interaction (such as silicon), the spin relaxation times can turn out to be short.

In this work, we report on the effect of spin-dependent scattering in a system of free and donor-bound electrons in epitaxial GaAs films doped with Si. Spin-dependent scattering manifests itself in the appearance of magnetoresistance in weak fields (H~1 G). Under conditions of optical orientation of electron spins, the application of an external magnetic field along the direction of the electric current passed through the sample leads to a decrease in the electric resistance of the material. The proposed explanation of the effect is a decrease in the efficiency of spin-dependent scattering of initially spin-oriented electrons in a magnetic field caused by spin depolarization of electrons. The assumption is confirmed by experiments on the optical orientation of spins and magnetic depolarization, i.e. the Hanle effect.

Interest in gallium arsenide in the context of spin-dependent transport is due to both the wide range of applications of the material in modern electronics and well-studied spin properties. However, the spin transport in gallium arsenide, in contrast to silicon, has not been studied in sufficient detail. Apparently, this is because of the fact that the spin relaxation times in GaAs are much shorter than in Si (spin–orbit interaction in GaAs is much stronger than in Si), which does not allow usage of electron paramagnetic resonance methods to study the spin properties of nonequilibrium carriers.

**Experiment.**

The gallium arsenide epitaxial film with a thickness of 1.5 μm and a donor (Si) concentration of $1.15 \times 10^{16}$ cm$^{-3}$ was studied. The choice of the sample was due to the relatively long spin lifetimes of electrons, exceeding 300 ns, which are observed in bulk gallium arsenide doped with donors [17].

In the study of spin-dependent phenomena, either optical or transport methods are usually used. We combine these methods for the most reliable interpretation of experimental data. Another feature of our experiments is that the magnetic field was applied along the current, so that it does not affect the motion of carriers (the Lorentz force is excluded), but it does affect the spin polarization of electrons (the latter can be observed simultaneously in the optics by the PL polarization measurements). In addition, in the papers [8,10,11,13], the spin polarization was created by a strong magnetic field, and therefore it is difficult to separate the possible contributions to the magnetoresistance. In our experiments, the spin polarization was created optically, and the effect was registered by depolarization in a magnetic field of 1 G.

To generate spin-polarized carriers, we used the optical orientation method. The carriers were excited by a circularly polarized ($\sigma^+$) beam of a Ti:sapphire laser. The sample was placed inside a cryostat with a variable temperature insert that allows one to vary the temperature from 1.4 to 300 K. The degree of circular polarization of the PL $\rho_c = (I_+ - I_-)/(I_+ + I_-)$ was measured, where $I_+$ and $I_-$ stand for the intensities of right ($\sigma^+$) and left ($\sigma^-$) circular polarizations. Magnetooptical measurements were made in Voigt geometry.

The magnetotransport properties of spin-polarized electrons were studied using the 4-contact method. A constant voltage was applied to the sample, and the current flowing through the sample was measured by an ammeter (see inset in Fig. 2). A voltage signal was taken from the measuring contacts, which made it possible to determine the resistance of the sample. In order to separate the contribution associated with photoexcited spin-oriented carriers, synchronous detection was used. The laser intensity was modulated by a chopper at a frequency of 183 Hz, while the signal from the measuring contacts was demodulated using a synchronous detector (lock-in). The voltage measured by a synchronous detector is proportional to the difference in the sample resistance in the dark and in the light.

Figure 1 shows the intensity and the degree of circular polarization spectra of the PL. The most intense line at a photon energy of E=1.5145 eV corresponds to an exciton localized on a neutral donor (D$^0$X) [18]. The highest degree of circular polarization corresponds to the line of free excitons (E=1.5158 eV [18]). Electrons photoexcited by circularly polarized light thermalize to the bottom of the conduction band in a short time partially conserving their spin orientation. However, the radiation on the line of excitons localized on the donors is practically unpolarized, since the two electrons in trion complex is in the singlet state (holes in bulk semiconductors like GaAs are usually depolarized because of very short spin relaxation time). On the contrary, the free exciton line emission has a relatively high degree of circular polarization, which is directly determined by the state of the electron spin system.

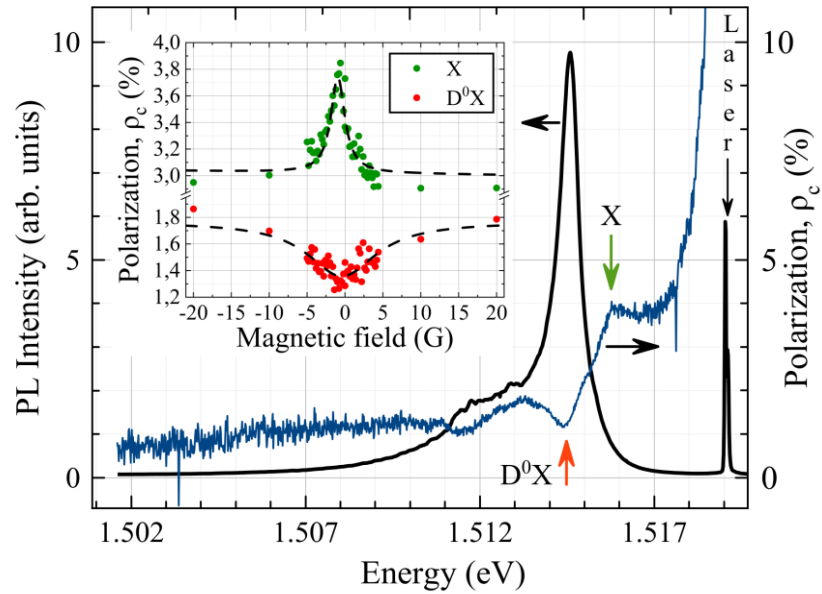

FIG. 1. Spectra of intensity and degree of circular polarization of PL excited by circularly polarized light with a photon energy of 1.519 eV. The optical excitation power was 2.8 W/cm$^2$. Temperature = 1.4 K. The inset shows the Hanle curves for the detection energies corresponding to free excitons (1.5158 eV, red symbols) and for donor bound excitons (1.5145 eV, green symbols). Experimental curves are fitted by Lorentz contours (dashed lines).

The magnetic field was applied along the surface of the sample, parallel to the current and perpendicular to the light-induced spin polarization. The results of the magnetoresistance experiments are shown in Fig. 2. With circularly polarized illumination, the voltage measured by the synchronous detector demonstrates a dependence on the magnitude of the magnetic field. The resulting sample resistance decreases while illuminated due to extra carriers photogeneration, thus the $\Delta R$, which is the difference between the resistance of the sample in the in the light and in the dark, is negative. The application of a magnetic field causes a further decrease in the resistance, $\Delta R$, see figure 2. The effect does not depend on changing the sign of the light circular polarization and can be described by a Lorentz contour. Note that when excited by linearly polarized light, the magnetoresistance effect was not observed, which indicates the spin nature of the observed phenomenon. The change in resistance occurs in small fields, the width of the Lorentz contour is 1 G, i.e. is close to the width of the Hanle effect measured on the line of free excitons (2 G), see inset in Fig. 1. It can be concluded that both the Hanle effect and the magnetoresistance effect are determined by the same physical process - the depolarization of electron spins in the magnetic field. Let's fix the result - the magnetic field leads to depolarization of electrons and, as a consequence, to a decrease in the resistance of the sample.

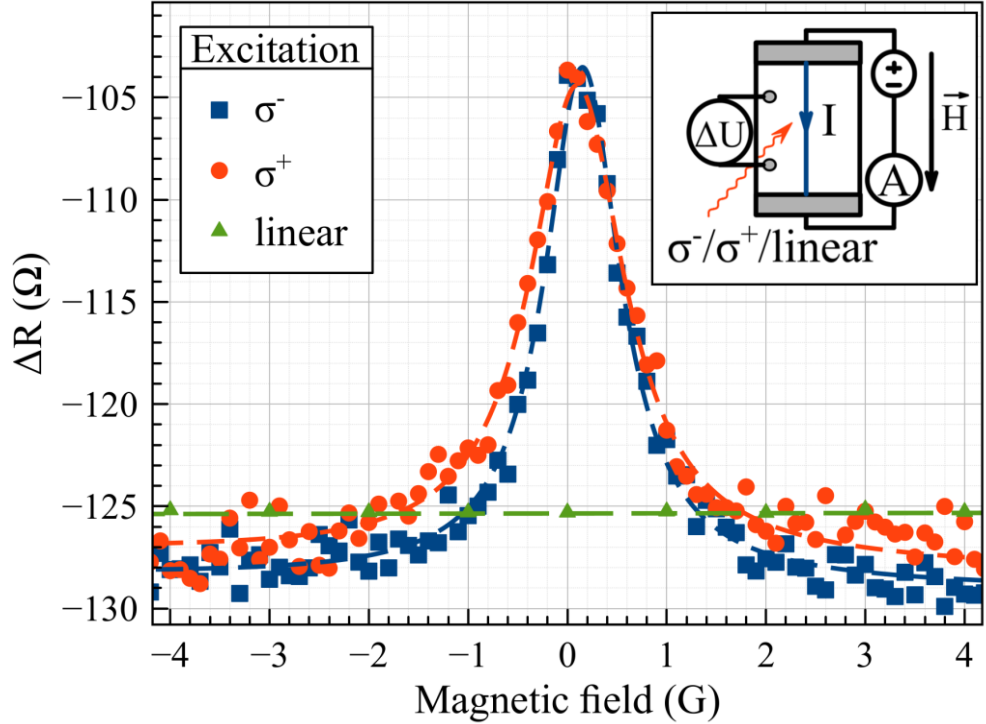

FIG. 2. The difference between the resistance of the sample when illuminated by a laser beam and the resistance in the absence of illumination. The experimental data obtained under left circular polarized excitation are shown by blue squares, the corresponding approximation blue dashed line is a Lorentz contour. Similarly, the red circles and the dashed line correspond to an excitation with right circular polarization. Green triangles and the dashed line correspond to the case of linearly polarized excitation. The current flowing through the sample was I = 217 µA, the optical excitation power was 0.65 W/cm$^2$, and the excitation photon energy was $E_{exc}$ = 1.518 eV. The inset shows a diagram of the electrical measurements. Temperature = 1.4 K.

The important feature of the observed magnetoresistance effect is its parity with respect to the sign of the circularly polarized excitation ($\sigma^+/\sigma^-$). The parity apparently indicates the simultaneous participation of spin-oriented free and localized carriers, so that the effect is determined by the product of their polarizations. A more detailed discussion is given in the following section.

The amplitude of the effect $\delta_H = \Delta R(0) - \Delta R(\infty)$, where $\infty$ corresponds to the saturation field of the dependence plotted on Fig. 2, characterize the change in electrical resistance in a magnetic field. $\delta_H$ depends on the energy of photons of the exciting radiation, temperature and strength of the current passed through the sample. Figure 3 shows the dependence of the $\delta_H$ on the energy of exciting photons. The dependence has a resonant character with a maximum corresponding to the energy of free excitons. This dependence is caused by the fact that when the excitation photon energy is greater than the exciton energy, the polarization of carriers is usually smaller than in the case of resonant excitation of exciton [19]. At excitation energies lower than the exciton energy, the carriers are simply not excited, and the effect disappears.

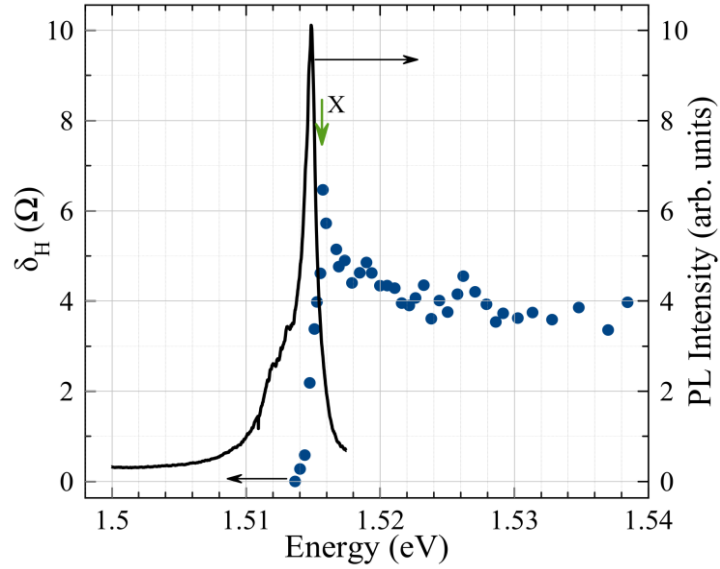

FIG. 3. The change in electric resistance $\delta_H$ as a function of the energy of the exciting photons (blue circles). The photoluminescence spectrum is shown by the solid black line. A current of I = 217 µA was passed through the sample, the power density of the exciting laser was 0.16 W/cm². Temperature = 1.4 K

Figure 4(a) shows the magnetoresistance curves for two values of the current. As one can see, as the current increases, the value $\delta_H$ goes to zero. The range of current values, where the amplitude of the effect is the largest from 0 to 2V. The highest amplitude of the effect corresponds to a voltage value of 1.9 V. With a further increase in voltage, the avalanche growth of the current takes place due to the knocking out of carriers from donors (inset in Fig. 4(b)). At current values exceeding 800 µA, the effect is not observed.

We obtained a temperature dependence of the value $\delta_H$ (Fig. 4(a)). With increasing temperature, the value of the $\delta_H$ decreases, the effect is not observed at temperatures exceeding 5K. Figure 4(c) shows the dependence of the absolute value of the resistance of the sample under the condition of optical pumping on the magnitude of the magnetic field in the Voigt geometry. In a zero magnetic field, the sample has a resistance of 6261 Ω, with an increase in the field amplitude, the resistance drops to 6250 Ω (change by 0.16%).

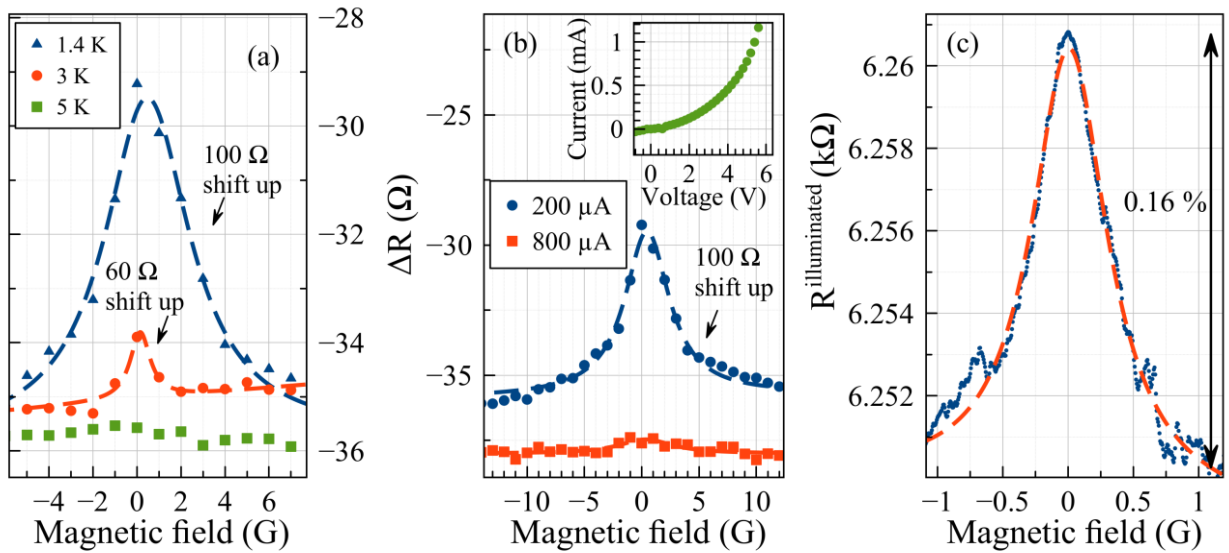

FIG. 4. (a) Difference in resistance in the dark and in the light subject to magnetic field at temperatures of 1.5, 3 and 5K. The power density and laser excitation energy are 0.16 W/cm² and

1.518 eV, respectively. Values for 1.4 K and 3 K are shifted for ease of comparison by 100 Ω and 60 Ω, respectively. (b) Difference in resistance for current values of 200 µA (blue circles and dashed line) and 800 µA (red squares) subject to magnetic field. Values for 200 µA are shifted by 100 Ω for ease of comparison. Temperature = 1.4 K. The inset shows the current–voltage characteristic of the sample. (c) Dependence of the absolute sample resistance subject to magnetic field. The current value is 219 µA. The power density and excitation energy of the laser are 0,16 W/cm$^2$ and 1.518 eV, respectively. Temperature = 1.4 K

**Discussion.**

One of the possible mechanisms for changing the photoconductivity is the spin-dependent capture of charge carriers. To exclude the assumption, that observed here effect of magnetoresistance is due to the spin-dependent capture, it is sufficient to make sure that the intensities of neutral and donor-bound excitons photoluminescence do not change when the electron spins are depolarized by a magnetic field. Spin-dependent scattering does not contribute to the PL intensity, while spin-dependent trapping should lead to a change in the PL intensity. However, we did not register any changes in the intensity of PL in a magnetic field, so it is unlikely, that the spin-dependent capture is essential here.

As an interpretation of the obtained magnetoresistance dependences, we can offer the following reasoning. Since the effect is observed only when carriers are excited by circularly polarized light, it is obvious that the phenomenon is caused by the spin polarization of charge carriers. This is also confirmed by the resonant nature of the dependence of the effect amplitude on the energy of the exciting laser (the maximum of the effect corresponds to the energy of free excitons).

In addition, the parity of the effect with respect to the polarization of the laser beam ($\sigma^+$/$\sigma^-$) indicates the existence of correlation in the system of optically oriented free electrons and electrons localized on donors. The fact that the values of the Hanle effect width and the width of the dependence of the resistance on the magnetic field (2 and 1 G) are close (the difference by a factor of two can be explained by different pump densities) confirms this correlation, since the first one (the Hanle effect width) is extracted from data on the polarization of the emission of free excitons, while the second is determined by the polarization of free and localized carriers. Thus, we believe that both spin-oriented free and localized on donors electrons participate in the formation of the magnetoresistance signal.

In accordance with the foregoing, the model of the phenomenon is reduced to the problem of the scattering of polarized electrons by a polarized target (in our case it is neutral donor) (Møller scattering [20]). Due to the previously mentioned exchange mechanism, the spins of free and localized electrons are averaged after photoexcitation. When a transverse (with respect to the spin) magnetic field is applied, free and captured electrons on donors are depolarized, which changes the scattering rate, thereby changing the mobility of carriers and, at the same time, the resistance of the sample.

Additional evidence in favor of the proposed mechanism is a decrease in the amplitude of the effect with an increase in temperature or the current passed through the sample. Since in both cases the number of localized carriers decreases, the rate of exchange (spin-dependent) scattering of free carriers will decrease and, finally, when the donors are completely depleted, the effect will disappear in accordance with the experiment.

To quantify the observed effect, we use the analogy between scattering by a neutral impurity in a semiconductor and the scattering of slow electrons by a hydrogen atom. The corresponding theory describes scattering in terms of singlet *s* and triplet *t* contributions, and differential scattering cross section is given by [21, 22]:

$$\frac{d\sigma}{d\omega} = s\left(|f+g|^2\right) + t\left(|f-g|^2\right), \tag{1}$$

Here, *s* and *t* give the fraction of singlet (antiparallel spins of the incident and scattering electron) and triplet (parallel spins) collisions, *f* and *g* are scattering amplitudes corresponding to the absence of exchange and the presence of particle exchange during electron scattering. *f* and *g* - appear in the processes of singlet and triplet scattering due to the requirement of antisymmetry on the wave functions.

For an arbitrary spin polarization *P* of incident and scattering electrons, one can determine the total scattering cross section as [21, 22]:

$$\sigma = \frac{(1-P_e P_d)}{4}\int_{4\pi}|f+g|^2 d\omega + \frac{(3+P_e P_d)}{4}\int_{4\pi}|f-g|^2 d\omega, \qquad (2)$$

where $\frac{(1-P_e P_d)}{4}$ and $\frac{(3+P_e P_d)}{4}$ are the fractions of singlet and triplet states correspondingly.

Using the values of the constants for the hydrogen atom and assuming that such a consideration is valid for the case of scattering by hydrogen-like centers in a semiconductor (taking into account the scaling factors - effective mass, dielectric constant), one can write [28]:

$$\sigma = C(54.3 - 32.6 P_e P_d), \qquad (3)$$

From this equation it can be seen that the electrical resistance of the sample, *R*, can both increase and decrease with an increase in the polarization of carriers, depending on the mutual direction of the spins of localized and free electrons [22]. With the same signs of polarizations, the electrical resistance $R \sim \sigma$ increases with increasing magnetic field (depolarization of carriers), and with different signs of polarizations, it decreases. The latter case (different signs of polarizations) corresponds to the sign of the change in the sample resistance in our experiments. Let us take the value of the degree of electron polarization P=0.05, estimated from optical experiments (X line on Fig.1), and assume that the polarization of electrons localized on donors has the same absolute value [23]. The exchange interaction over short times (on the order of ten picoseconds) compared to the carriers' lifetime leads to an averaging of the degree of spin polarization between the two subsystems. Then, in the case of different signs of the polarization of the spins of free and donor electrons, the corresponding ratio of the scattering cross sections will be equal to 1.0015, i.e., the change in resistivity will be 0.15 %. The value of the change in resistance obtained in this way agrees well with the experimental value obtained during the depolarization of electrons in a magnetic field and is 0.17 % (see Fig. 5). Note that the good agreement between our estimate of the change in resistance and the theoretical value means that scattering by a neutral impurity is the dominant carrier scattering mechanism in the sample. However, if we assume that the signs of the polarizations of the spins of free and donor electrons are the same, as in the case studied by Page [23], then we obtain the opposite sign of the magnetoresistance effect, which contradicts the experiment. Thus, the model explains the experiment under the condition that the signs of the polarization of the spins of free and donor electrons are opposite. Perhaps the Paget model developed for moderate donor concentrations ($N_D<10^{14}$ cm$^{-3}$) does not work in the current case, the case of high donor concentrations (donor concentration $N_D \sim 10^{16}$ cm$^{-3}$ corresponds to the metal-insulator transition in n-GaAs). An argument in favor of the opposite direction of the spins of free and localized electrons is the Hanle curve on the $D^0 X$ line. The curve has a reversed form, the polarization increases with magnetic field (see the caption in Fig. 1), what can be interpreted as a decrease in the contribution with a negative degree of polarization against the background of a constant positive contribution, a similar situation was discussed in Ref. [24]. Negative polarization may come from a recombining with a hole donor bound electron with a spin flipped with respect to the optical pumping orientation.

**Conclusion.**

In this work, we studied the effect of magnetoresistance in bulk samples of n-type gallium arsenide (donor concentration of $1.15 \times 10^{16}$ cm$^{-3}$) in extremely weak magnetic fields of ~1 G. It has been experimentally shown that the effect is related to the spin polarization of charge carriers. All magnetoresistance effects known from the literature occur in much higher fields. Often, a magnetic field is used to create spin polarization of charge carriers. In our experiments, spin polarization was created using optical pumping, and a weak magnetic field applied perpendicular to the spin led to the depolarization of electron spins and, as a consequence, to a change in a resistance of the sample. This phenomenon can be considered as an electrical registration of the Hanle effect, known from optics since 1924 [25]. To confirm the interpretation of the results, we measured the optical orientation of the spins in terms of the polarization of the radiation and the depolarization in a transverse magnetic field. The results of optical and magnetotransport measurements are in good agreement with each other. The most probable reason for the decrease in the signal of magnetoresistance is the change of the singlet and triplet contributions to the cross section for the scattering of band electrons by neutral donors in the case of spin depolarization by a transverse magnetic field. This model makes it possible to achieve good agreement with the observed effect, predicting a change in resistance of 0.15 % at an observed value of about 0.17 %. At the same time, the details of the interaction between the spins of free and localized electrons remain unclear. In particular, we cannot say what process is responsible for the different signs of polarizations of the spins of free and localized electrons. These questions are the subject of further research.

**Acknowledgement.**
All activities conducted by M. D. Ragoza, N. V. Kozyrev, S. V. Nekrasov, B. R. Namozov, and Yu. G. Kusrayev, including optical and electrical measurements, were supported by the RSF (project No. 23-12-00205). The work of N. Bart, A. Ludwig, and A. D. Wieck was supported by the DFG TRR160.

**References.**

[1] M. I. Dyakonov, *Spin physics in semiconductors*, (Springer, Berlin, 2017).
[2] A. Fert, Rev. Mod. Phys. **80**, 1517 (2008).
[3] D. D. Awschalom, D. Loss, and N. Samarth, *Semiconductor spintronics and quantum computation*, (Springer, Berlin, 2002).
[4] Z. H. Xiong, D. Wu, Z. V. Vardeny, and J. Shi, Nature **427**, 821 (2004).
[5] G. Bergmann, Int. J. Mod. Phys. B **24**, 2015 (2010).
[6] B. L. Al'tshuler, A. G. Aronov, A. I. Larkin, and D. E. Khmel'nitskil, Zh. Eksp. Teor. Fiz. **81**, 768 (1981) [Sov. Phys. JETP **54**, 411 (1981)].
[7] G. D. Mahan, Many-Particle Physics, (Plenum, New York, 2000), 3rd ed.
[8] C. Erginsoy, Phys. Rev. **79**, 1013 (1950).
[9] H. S. W. Massey and B. L. Moiseiwitsch, Phys. Rev. **78**, 180 (1950).
[10] R. Maxwell and A. Honig, Phys. Rev. Lett. **17**, 188 (1966).
[11] L. Qing, J. Li, I. Appelbaum, and H. Dery, Spin relaxation via exchange with donor impurity-bound electrons, Phys. Rev. B **91**, 241405(R) (2015).
[12] G.L. Bir, A.G. Aronov, G.E. Pikus, Zh. Exp. Teor. Fiz. **69**, 1382 (1975) [Sov. Phys. JETP **42**, 705 (1976).]
[13] Y. Yafet, in Solid State Physics, edited by F. Seitz and D. Turnbull (Academic, New York, 1963)
[14] R. Elliott, Phys. Rev. **96**, 266 (1954).
[15] M. I. D'yakonov and V. I. Perel', Sov. Phys. Solid State **13**, 3023 (1972).
[16] M. I. D'yakonov and V. I. Perel', Zh. Eksp. Teor. Fiz. **60**, 1954 (1971). [Sov. Phys. JETP **33**, 1053 (1971)].

[17] R. I. Dzhioev, B. P. Zakharchenya, V. L. Korenev, et al., Pis'ma Zh. Éksp. Teor. Fiz. **74**, 200 (2001). [JETP Lett. **74**, 182 (2001).]
[18] E. H. Bogardus and H. B. Bebb, Phys. Rev. **176**, 993 (1968).
[19] F. Meier and B. Zakharchenya, *Optical Orientation* (North Holland, Amsterdam, 1984), Vol 8.
[20] J. Kessler, *Polarized Electrons*, (Springer, Berlin, 1985), 2nd ed.
[21] A. Honig, Phys. Rev. Lett. **17**, 186 (1966).
[22] R. Maxwell and A. Honig, Phys. Rev. Lett. **17**, 188 (1966).
[23] D. Paget, Phys. Rev. B **24**, 3776 (1981); **25**, 4444 (1982).
[24] R. I. Dzhioev, B. P. Zakharchenya, V. L. Korenev, P. E. Pak, D. A. Vinokurov, O. V. Kovalenkov and I. S. Tarasov, Phys. Solid State **40**, 1587 (1998)
[25] W. Hanle, Z. Phys. **30**, 93 (1924).